\documentclass[12pt,preprint]{aastex63}

\usepackage{lineno}

\shorttitle{Sirius on 2MASS System}
\shortauthors{Rieke et al.}

\begin{document}

\title{Infrared Absolute Calibration I: Comparison of Sirius with Fainter Calibration Stars}
\author{G. H. Rieke}
\affiliation{Steward Observatory, University of Arizona, Tucson, AZ 85721, USA, also Department of Planetary Sciences}
\author{Kate Su}
\affiliation{Steward Observatory, University of Arizona, Tucson, AZ 85721, USA}
\author{G. C. Sloan}
\affiliation{Space Telescope Science Institute, 3700 San Martin Drive, Baltimore, MD 21218, USA}
\affiliation{Department of Physics and Astronomy, University of North
  Carolina, Chapel Hill, NC 27599-3255, USA}
\author{E. Schlawin}
\affiliation{Steward Observatory, University of Arizona, Tucson, AZ 85721, USA}

\begin{abstract}

A challenge in absolute calibration is to relate very bright stars with physical flux measurements to faint ones within range of modern instruments, e.g. those on large groundbased telescopes or on the James Webb Space Telescope (JWST). We propose Sirius as the fiducial color standard: it is an A0V star that is slowly rotating and does not have infrared excesses either due to hot dust or a planetary debris disk;  it also has a number of accurate ($\sim$ 1 $-$ 2\%) absolute flux measurements.  We transfer the near infrared flux from Sirius  accurately to BD +60 1753, an unobscured early A-type star (A1V, V $\approx$ 9.6, E(B-V) $\approx$ 0.009) that
is faint enough to serve as a primary absolute flux calibrator for JWST.
Its near-infrared spectral energy distribution and that of Sirius should be virtually identical.  We have determined its output relative to that of Sirius in a number of different ways, all of which give consistent results within $\sim$ 1\%.  We also transfer the calibration to GSPC P330-E, a well calibrated close solar analog (G2V). 
We have emphasized the 2MASS $K_S$ band since it represents a large number and long history of measurements, but theoretical spectra (i.e., from CALSPEC) of these stars can be used to extend this result throughout the near- and mid-infrared. 

\end{abstract}


\section{Introduction}

Very accurate physical calibration is desired in a number of fields of astronomy, as summarized in \citet{kent2009}. The highest accuracies are currently achieved in the visible, with nominal errors of order 0.5 - 1\%  at 5556 \AA ~\citep[e.g.,][]{bohlin2014a}. Nonetheless, work continues  in that spectral range \citep[e.g.,][]{bohlin2014, bohlin2017, deustua2018}, and achieving a robust 1\%-quality calibration that can be applied to routine photometry is very challenging (see, e.g., \citet{bohlin2019}). White dwarf stars have proven valuable to provide spectral templates  to extend the calibration spectrally over the entire visible and ultraviolet range \citep[e.g.,][]{bohlin2019, bohlin2020}. There is increasing interest in extending use of white dwarf spectral templates into the infrared \citep[e.g.,][]{bohlin2011, fusillo2020}, which can provide the necessary tests of their spectral models in the mid-IR. However, they become too faint for routine use at the longer infrared wavelengths with most instruments. In addition, many of them may have infrared excess emission due to reradiation of the white dwarf emission by surrounding dust \citep{farihi2016}, a source of flux that is, of course, not included in the extrapolated spectral templates. Also, the most fundamental absolute calibrations trace directly to calibrated reference sources and hence are generally confined to the brightest stars; white dwarfs are far out of reach for such measurements.  

Infrared calibration grew from the main sequence stellar calibration introduced by \citet{johnson1953}, which defined zero magnitude (just for the U, B, and V bands) on the basis of a suite of A0V stars. Particularly in the mid-infrared, the original goal of using multiple stars was impractical and the system has been strongly based on Vega, both because it was readily measured even through the Q band at 20 $\mu$m, and because it was bright enough for direct calibration experiments using reference blackbody sources. Indeed, the resulting photometric system is commonly described as the ``Vega'' system. Unfortunately, Vega is not suitable as the defining star for a system at the accuracy levels we now aspire to. Its shortcomings are legend. First, it has a huge far infrared excess \citep{aumann1984} that persists at a measureable level down through 10 $\mu$m \citep{su2013}.  Second, the star is a pole-on rapid rotator with a significant temperature gradient from pole to equator \citep{aufdenberg2006}; in fact, rather than defining zero color, the $V - K$ color of the stellar photosphere is $\sim$ 0.045 \citep{rieke2008}. Third, Vega has an excess of $1 - 2$\% in the near infrared above its photospheric emission \citep{absil2006}, possibly due to hot dust confined magnetically \citep{rieke2016}, which might even be variable \citep{ertel2016}. 

Nonetheless, there is by now a huge body of photometry based on the A0V star zero point, which any new approach must preserve. Fortunately, Sirius may in fact serve as an ``ideal'' A-star to define a much better-understood zero point, both as a metric for models of stellar atmospheres and as a photometric reference. Sirius is of spectral type A0mA1Va \citep{gray2003}, which means that it is a mild Am star. The classification indicates that the star is A0V in type judged by its hydrogen lines (the ``a'' means it is of high luminosity for the main sequence) but A1V when judged by lines of metals. This is different from saying that its spectral type is between A0 and A1; it is an A0 star, but has line strengths like that of an A1 star, meaning that it is somewhat metal-rich. Both its anomalous abundances and its low vsini $\sim$ 16 km s$^{-1}$ suggest that it rotates relatively slowly \citep{royer2002,  mathys2004, gray2014, takeda2020}. It appears to have little or no excess due to a debris disk \citep[e.g.,][]{su2006}, nor does it show a hot excess near 2 $\mu$m \citep{kervella2003}. Since Sirius is a ``model'' A0V star, its spectral energy distribution is well-represented by theoretical models \citep[e.g.,][]{bohlin2014,bohlin2017}. Sirius also has a number of accurate measurements of its absolute flux, most notably by the Mid-Course Space Experiment (MSX) mission \citep{price1999}. There have been a number of previous suggestions to switch to Sirius as the primary defining star for stellar photometry \citep[e.g.,][]{engelke2010,bohlin2014, krisciunas2017}. This paper takes a step toward completing this goal. 

The paper is the first in a series in which we address the issues to establish an accurate absolutely calibrated infrared photometric system  that is based on Sirius and also is applicable down to standard stars at faint magnitudes. The challenges in doing so are: (1) to tie the huge output of Sirius (and other absolutely calibrated stars) accurately to much fainter stars that are useful for calibration with modern detector systems (Sirius saturates virtually any modern system in the optical and infrared); and (2) to calibrate the resulting revised photometric system in physical units (to be discussed in following papers). 

The 2MASS \citep{skrutskie2006} photometric system covers the entire sky uniformly \citep{cutri2003}  and has become the default system for near infrared photometry \citep[e.g.,][]{warren2007, hodgkin2009}. We focus on the 2MASS $K_S$ photometric band because: (1) it is at a wavelength where stellar spectral energy distributions (SEDs) are relatively free of strong absorptions that complicate synthetic photometry; (2) it is far enough into the infrared that extinction is small for the nearby stars we consider; (3)  it is easily connected to longer infrared wavelengths since the stars we consider have roughly Rayleigh-Jeans SEDs beyond $K_S$ with modest deviations due to spectrally localized absorptions; and (4) there is a large database of measurements in this band or similar ones. The goal of this paper is to derive accurately the brightness of Sirius in the $K_S$ band relative to the much fainter stars that are accessible to photometry by modern instruments. 
  Extending  the recalibrated system to the shorter infrared bands, e.g., $J$ and $H$, is reserved for future work, since it requires the establishment of a suitable zero point standard star - the subject of this paper. In addition, achieving the same level of accuracy in these bands is made more complicated by the larger interstellar extinction corrections.   
  
We also focus on three stars that have been used extensively in previous calibration efforts. First, BD +60 1753 is a fainter Sirius clone at spectral type A1V. It has a $K_S$ magnitude of 9.6,  within the range of a number of modes for all the instruments on JWST as well as being sufficiently faint for measurement with near infrared imagers on most large telescopes using narrowband filters. It has more than 400 measurements with Spitzer \citep{werner2004,gehrz2007} in the IRAC \citep{fazio2004} bands 1 and 2, and nearly 100 in Bands 3 and 4. Second, we use HD 165459 as a transfer standard; it is of type A3V and with more than 7000 Spitzer measurements in IRAC Bands 1 and 2. It is at $K_S \sim 6.6$, providing an intermediate step toward the fainter standards, important for the direct comparison with Sirius \citep{su2021} and for comparison with many of the A-stars with IRAC measurements in \citet{krick2021}, which are of similar brightness\footnote{Beyond $\sim$ 10 $\mu$m, HD 165459 has a significant debris disk excess \citep{sloan2015}; however, this does not compromise its use as a transfer standard at wavelengths $\le$ 5 $\mu$m.}. The use of both HD 165459 and BD +60 1753 as IRAC standards allows a direct comparison to transfer measurements to the latter star at high accuracy. Third, GSPC P330-E is a G2V star that is a close solar analog and was a prime calibrator for NICMOS \citep{thompson1998} on HST \citep{odell1994}. At a magnitude of $K_S \sim 11.4$, it is even more accessible to imagers on large groundbased telescopes.  All three of these stars have been very extensively characterized, particularly in the visible range, with results available in the CALSPEC database \citep{bohlinrev14, bohlin2017}. We remark, however, that steps toward even fainter standard star systems have proven desirable for some applications \citep{leggett2020}. Additional effort will also be necessary to extend the calibration at full accuracy to the other near infrared bands (e.g, there are $\sim$ 2\% offsets from zero colors for A0V stars in 2MASS $J - H$ and $H - K_S$ \citep{rieke2008, maiz2018}). 

We proceed as follows. In \S 2, we first introduce the Spitzer/IRAC measurements that are used in this paper to (1) establish the color-color relation of $V - K_S$ and $K_S -$ IRAC1 for main-sequence stars and (2) assess the stability of the photometry and determine a direct comparison of the routine calibrators BD+60 1753 and HD 165459 over the entire Spitzer mission. In \S 3, we derive the 2MASS-equivalent $K_S$ magnitude of Sirius using five distinct approaches, including that developed in the accompanying paper by \citet{su2021} that derives a direct transfer from Sirius to HD 165459 in the IRAC Bands 1 and 2. The agreement among all of these approaches is excellent, within $\pm$ 0.006 mag, indicating that systematic errors in any of them are small.  In \S 4, we extend the work to fainter stars, including BD +60 1753 (A1V at $K_S$ = 9.6) and GSPC P330-E (G2V at $K_S$ = 11.4).  \S 5 briefly discusses the challenges in establishing an accurately calibrated network over the entire sky. The major results are summarized in \S 6. In addition, there are three appendices.   \S A describes the transformations from heritage photometric systems into the 2MASS $K_S$ band. \S B derives the color term between $K_S$ and IRAC Band 1 as a function of $V-K_S$ and also provides a test of 2MASS $K_S$ magnitudes for moderately bright stars transformed from heritage photometry. \S C tests the transformed $K_S$ magnitudes of very bright stars for accuracy by comparing them with the photometry from the Diffuse Infrared Background Experiment (DIRBE) instrument on the Cosmic Background Explorer (COBE) \citep{mather1990,silverberg1993}. 

\section{Spitzer/IRAC Data Processing}

Most of the measurements used in this paper were taken from published sources. However, we carried out our own reductions of the IRAC data we use to (1) determine the color correction between $K_S$ and IRAC Band 1 for main sequence stars; (2) determine a {\it direct} transfer between the  two Spitzer IRAC calibration stars HD 165459 and BD +60 1753, both as a test of the photometric stability and as the most accurate comparison possible between these stars; and (3) do a direct transfer from Sirius to fainter stars.

To determine the intrinsic color difference between $K_S$ and IRAC Band 1, we used the observations obtained in Spitzer PID 70076. We focus on Band 1 to avoid the effects of the strong CO absorption in Band 2 for stars later than mid-F-type. The data were processed by the Spitzer Science Center (SSC) with the final IRAC pipeline \citep{lowrance2016}. We used the BCD (basic calibrated data) images, which have a native scale of 1\farcs22 pixel$^{-1}$.  We performed aperture photometry using an on-source radius of 3 pixels and a sky annulus between 3 and 7 pixels,  with an aperture correction factor of 1.1233, as in \citet{krick2021}.  The BCD photometry was corrected for the pixel solid angle (i.e., distortion) effects based on the measured target positions using files provided by the SSC. We also discarded any photometry when the target was too close to the edge of the detector array, and obtained weighted average photometry for each of the astronomical observation requests (AORs) by rejecting the highest and lowest photometry points in the same AOR. The same procedures were also conducted on the mosaic post-BCD products for comparison. A few photometry points from the post-BCD products were significantly fainter (up to 10\%) than the values obtained from the individual BCD images. Inspection of those AORs found that one or two BCD frames have poor WCS (world coordinate system) information, likely causing misalignment in the post-BCD products. Thus, we adopted our final photometry based on the weighted average of the BCD images, which were not subject to this issue. 

We also made a high weight {\it direct} comparison between BD +60 1753 and HD 165459 based on the huge number of observations of each of these stars as primary IRAC calibrators. Since they are of similar spectral type and low extinction, we can compare the signals from the two stars directly to obtain an accurate difference.  We used an automated pipeline to compute the IRAC photometry of the two stars throughout the cold mission and through August 2018 in the warm mission. This pipeline reduces all of the observations of these stars in a uniform way, using aperture photometry. The initial steps were similar to those described in the preceding paragraph. The signal was extracted with an aperture on the source of 10 pixels radius, relative to a sky annulus of inner radius 12 pixels and outer radius 20 pixels. These parameters were selected after experiments to determine the values giving the most reliable and accurate photometry.  Outliers were rejected per AOR by discarding the maximum and minimum flux values if the spread of the distribution was greater than 2\% and the AOR had at least 10 files.

Finally, we make use of IRAC photometry of very bright stars based on the wings of the saturated point spread functions. This work is described in detail in \citet{su2021}.

\section{The $K_S$ magnitude of Sirius}

Figure 1 provides an overview of the steps we have implemented to relate Sirius accurately to fainter stars in the $K_S$ band. The extreme left column shows infrared magnitudes progressing from nearly -2 for Sirius to fainter than 11 for GSPC P330-E. The next column to the right indicates the various stars and stellar samples that we use as reference standards. The double ended arrows connecting them indicate the existing photometry that provides measurements of their relative brightnesses. The center two columns show the various steps taken to relate these references to Sirius and the other target stars. The rightmost column includes the target stars and also some of the processing steps we have used to relate them directly to each other without passing through the details of a photometric system. As shown in Figure 1, we depend heavily on heritage $K$ photometry transformed into 2MASS $K_S$.  Details of the individual transformations are given in Appendix {\bf A}. By construction, such transformations should not introduce substantial biases. Specifically, in Appendix {\bf B}, we show that any bias is $0.0014 \pm 0.0021$ magnitudes, i.e., negligible. In Appendix {\bf C}, we utilize DIRBE photometry for comparison on bright stars, finding that the transformed $K_S$ magnitudes are generally accurate to $\sim$ 2\%, similar to the accuracy of 2MASS itself for sufficiently bright but unsaturated stars \citep{skrutskie2006}. Appendix {\bf B} finds a similar result for fainter stars. 

\begin{figure*}
\epsscale{.8}
\plotone{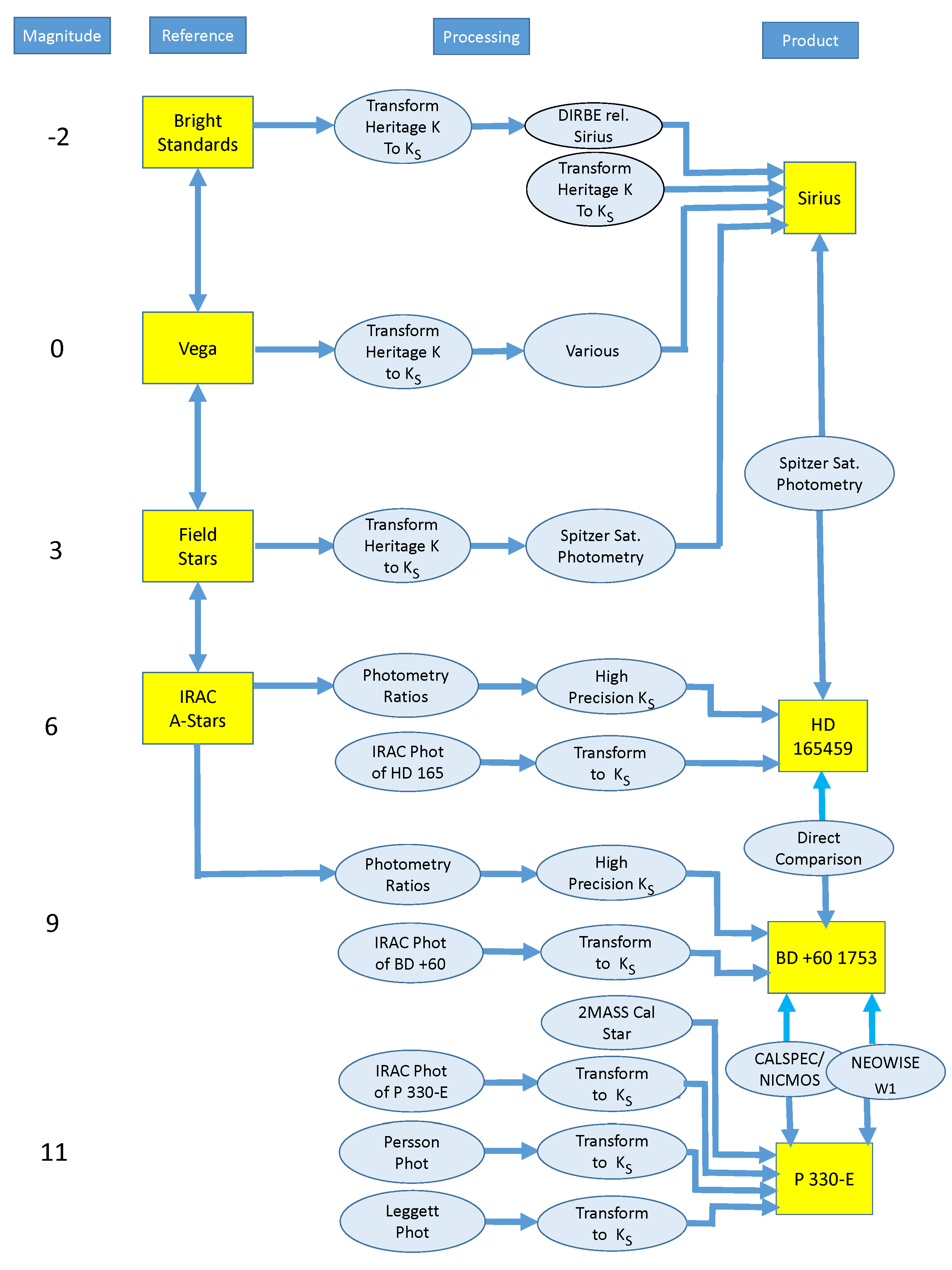}
\caption{Overall approach to calibration of Sirius at $K_S$ and to transfer this calibration to fainter stars relative to Sirius. The leftmost column shows fiducial magnitudes at K. The first column to the right of this one shows the starting points for the photometry, while the central two columns identify the steps we have used to put the outputs of these starting points onto a common photometric system. Where we started from photometry of the star to be calibrated, this is also registered in the center columns. The rightmost column shows the calibrated sources. Where additional calibration steps were taken among these sources, they are indicated within this column. }
\label{flow}
\end{figure*}

\subsection{Determination of Sirius $K_S$ with Traditional Photometry}

The most straightforward approach to calibrating Sirius is to use heritage photometry of the star, row 2 of Figure 1. Before standard infrared photometers lost the ability because they no longer had sufficient dynamic range, there were a number of measurements of Sirius. To make use of this photometry, we need to: (1) discard any measurements of questionable quality; and (2) put the remaining ones on the 2MASS photometric system. We only consider observing programs that addressed measurements of a significant number of stars {\it per se}\footnote{Studies using  Sirius as a calibration star in a program addressing some scientific priority other than photometry of large numbers of normal stars were rejected, since the heritage of the Sirius photometry for the rejected programs is difficult to determine and usually leads back to one of the accepted programs. }. 

The programs we have considered further are \citet{johnson1966,glass1974, engels1981, alonso1994, vdbliek1996} and the UKIRT  (United Kingdom Infrared Telescope)  bright standard star list \citep{ukirt2018}. We found, as discussed below, that consistent results for Sirius were obtained for the first five of these programs, but not for the UKIRT standards\footnote{Sirius is marked as ``a Mauna Kea (primary standard); JHKLL$'$M magnitudes for these objects should be accurate to $\pm$0.01 mag.''  \citet{leggett2003} gives a general description for the selection of the UKIRT standards, but she does not provide details to review the values for a specific star. However, we found Sirius in a listing of bright standards for the Infrared Telescope Facility (IRTF) at  \url{http://irtfweb.ifa.hawaii.edu/IRrefdata/Catalogs/bright\_standards}. The values are identical to those in the UKIRT list but they have a rather different notation: ``CIT (unpublished) ...  SJ sources; lower accuracy.  Avoid use.'' Combined with our finding that the Sirius value on this list is an outlier, we have adequate cause to discard it.}. As discussed further in the footnote, we have rejected the UKIRT value for Sirius from further consideration.

These programs are in differing photometric systems, since they were in an era when there was no completely uniform such system.  We therefore need to transform the $K$-band measurements from them accurately to the 2MASS photometric system. Doing so achieves two goals: (1) it reconciles the measurement frameworks by putting all the measurements on a consistent calibration; and (2) by adopting 2MASS $K_S$ for this calibration, it relates the brightness of Sirius to those of all the stars measured in the 2MASS survey.   Accurate transformations can be derived in the infrared, where virtually all of the relevant calibration stars (typically late B-type through early M-type dwarfs) have spectral energy distributions past $\sim$ 1.5 $\mu$m that are close to Rayleigh Jeans behavior, with modest deviations at spectral features (e.g., CO bands).

The general transformation of all heritage photometry onto the 2MASS system is useful in many ways for our study, as will become apparent as we proceed. To achieve high accuracy, we need to be mindful that, early in the development of near infrared photometry, systems of standard stars may have been updated as measurements were improved without the changes being documented.  The measurements of \citet{johnson1966} are homogeneous and extensive, but the stars are too bright for useful overlap with 2MASS; Appendix A describes the derivation of the transformation for them. To obtain the best results for Sirius from \citet{glass1974} and \citet{engels1981}, we carried out transformations using only the stars in the specific reference along with Sirius. We adopted this precaution because, as an example, although \citet{carter1990} and \citet{koen2007} have derived high quality transformations for the South African system, it is likely that the standard star values utilized in the reference we use, \citet{glass1974}, had been updated in the ensuing time. As an example, \citet{carter1990} introduced a change in the zero point.  For the two remaining references, \citet{alonso1994,vdbliek1996}, the general transformations were in fact based on photometry contemporaneous with the measurement of Sirius although not in the same reference.  Given that the \citet{johnson1966} measurements have fairly large scatter, we averaged the results (in magnitudes) on Sirius for $J$ and $K$. The final results are summarized in Table~1.  Although this value, $-$1.394, is based on only five sets of measurements, they agree well and also agree with a large number of independent determinations to be discussed below. It is difficult to assign {\it a priori} errors to photometric transformations; we characterize the errors by the $rms$ scatter of the results.

\begin{deluxetable}{lc}
\label{photometry}
\tabletypesize{\footnotesize}
\tablecaption{Transformed Photometry of Sirius to 2MASS $K_S$ Magnitudes}
\tablewidth{0pt}
\tablehead{
\colhead {Reference} &
\colhead {$K_S$ (2MASS)}
} 
\startdata
\citet{johnson1966}  &  -1.376  \\
\citet{glass1974}  &  -1.374  \\
\citet{engels1981}  &  -1.400  \\
\citet{alonso1994}  &  -1.398  \\
\citet{vdbliek1996}  & -1.424  \\
\hline
mean (average)  &  -1.394  \\
rms scatter  &  0.020  \\
standard error of the mean  & 0.010 \\
\enddata
\end{deluxetable}

\subsection{Use of DIRBE All-Sky Photometry to transfer from Bright Standards}

We now discuss row 1 in Figure 1. As demonstrated in Appendix C, the DIRBE photometry is very accurate (errors no more than $\sim$ 1\%) if care is taken to apply it only to stars with no contamination within the DIRBE beam. The calibration of DIRBE was based on a then-contemporary model of Sirius, so it cannot be used directly to determine the flux from this star. However, we can use the DIRBE measurements in a relative sense to compare the fluxes from Sirius with those from other stars. There are five stars that meet the anti-contamination criterion for accurate DIRBE measurements and for which there are accurate near-infrared spectral templates from \citet{engelke2006} so the color corrections can be calculated as needed to transfer the DIRBE 2.2 $\mu$m relative measurements accurately to the 2MASS passband (see Table 2). The $K_S$ magnitudes for these stars are based on very high weight heritage photometry, e.g., their use as standards by \citet{johnson1966} and a similar use for all but $\gamma$ Dra at the UKIRT. That latter star has a high weight measurement from \citet{kidger2003}.  Appendix {\bf B} shows that our transformations of heritage photometry to $K_S$ are accurate to the 1\% level and given the high quality of the heritage photometry for these five stars, we believe that the $K_S$ values for each approach this level of accuracy. The results from transferring the $K_S$ magnitudes of these stars to Sirius via the DIRBE measurements are in Table~2; their average yields a value of $-$1.402 for the $K_S$ magnitude of Sirius.

\begin{deluxetable}{ccccccc}
\tablecaption{$K_S$ magnitude of Sirius Using DIRBE Photometric Transfer
\label{dirbemethod}}
\tablehead{
\colhead {Star} &
\colhead {alt name} &
\colhead {Spec. Type} &
\colhead {$K_S$ mag} &
\colhead {rel Sirius} &
\colhead{color correction} &
\colhead{Sirius mag}
} 
\startdata
$\alpha$ Tau  &  HR 1457  & K5III  &  -2.837  &  1.474   &  -0.018  &  -1.381  \\
$\alpha$ Aur  &  HR 1708   & G3III &  -1.807   &  0.428   &  -0.005  &  -1.384  \\
 $\beta$ Gem & HR 2990 & K0III  & -1.118  &  -0.279  &  -0.009  &  -1.397  \\
$\alpha$ Boo &  HR 5340  & K1.5III  & -3.014  &  1.628  &  -0.017  &  -1.403  \\
$\gamma$ Dra & HR 6705  & K5III & -1.334 & -0.079  &  -0.018  &   -1.413 \\
\hline
mean (average)  &  &    &  &  &  &   -1.402  \\
rms scatter  &  &  &  &  &  &  0.020  \\
standard error of the mean  & &  &  & &  & 0.010 \\
\enddata

\end{deluxetable}

\subsection{Sirius Magnitude via that of Vega}

Another approach is to relate Sirius to Vega, for which there are a number of accurate transfers, and to then use the flux determinations for Vega to determine the flux of Sirius, as indicated in row 3 of Figure 1. We proceed by (1) comparing the brightness of the two stars; (2) deriving the Vega flux; and (3) combining these results with some additional corrections to determine a 2MASS $K_S$ magnitude for Sirius. 

\subsubsection{Sirius relative to Vega}

To determine an accurate magnitude difference in the infrared between Vega and Sirius, we focus on wavelengths between 1.25 and 7$\mu$m.This range avoids at the short wavelengths the strongest effects of the temperature gradient over the surface of Vega and at the long wavelengths, the debris disk infrared excess beyond 7$\mu$m. Comparing two stars of very similar spectral type has the major advantage that the details of the photometric system become largely irrelevant, since they affect the results for both stars in very similar ways. This is particularly true here, since both stars have smooth, Rayleigh-Jeans spectra (the hot dust excess adds a small contribution above the photosphere of Vega, but it also appears to be roughly Rayleigh-Jeans). All of the high-quality comparisons of Vega and Sirius in the infrared that we could find (high stability space observatories or groundbased measurements carefully reviewed for quality) are listed in Table 3. Nearly all of the entries are already known to be high-quality photometry. As discussed in Appendix {\bf C}, the DIRBE photometry is also likely to be highly accurate since there are no stars within the DIRBE beam brighter than $\sim$ 1\% of either Sirius or Vega, and, coincidentally, the brightest nearby star for each target is nearly the same fraction of the target flux, so no adjustment is required. We boosted the signal to noise of the DIRBE measurements by taking a weighted average of the magnitude differences at 2.2, 3.5, and 4.9 $\mu$m.

In making this comparison, we need to take account of the temperature gradient over the surface of Vega, which results in small departures from a pure A0V color at 1 $-$ 2.5 $\mu$m \citep{rieke2008}. Further departures result from the hot excess emission in the near infrared  \citep{absil2006}. Small adjustments are needed at J, H, and K to compensate both of these effects. 
To determine these adjustments, we start with the intrinsic $H-K_S$ and $J-K_S$  for Vega (i.e., including the effect of the temperature gradient), taken as 0.0021 and 0.011, respectively, from \citet{bessell1988},  \citet{koornneef1983}, \citet{mamajek2018}, and \citet{kidger2003}. We assume that the hot excess adds 1.3\%  both at K and around 3.5 $\mu$m and adds 0.7\% at $H$ \citep{bohlin2014}.

 We also need to link the longer wavelengths out to $\sim$ 5 $\mu$m to $K_S$. We first determine the magnitude difference between $K_S$ and the bands near 3.5 $\mu$m (W1, L, and IRAC Band 1) (we expect all of these bands to have the same magnitude on an A0V-based system). 
To base the comparison on A stars that are not rapid rotators, we use the compilation of Am and Ap stars in \citet{renson2009}, which should be dominated by slowly rotating stars \citep{royer2002}. We fitted the values of $J-K_S$ from 2MASS vs. the quoted stellar temperatures \citep{mcdonald2012} with a fourth power polynomial and rejected all stars deviating from the fit by 0.07 magnitudes or more (30 of 499). This step rejected stars with anomalous $J$ or $K$ photometry, and any with strong extinction. We then calculated $K_S$ - W1, rejecting all cases with indicated errors in this color $>$ 0.1 mag ($\sim$ 20\%), and fitted a linear relation to $J - K_S$ vs. $K_S$ - W1. This fit gives us a zero point ($K_S - W1$  for $J - K_S$ = 0) of  0.022 mag. We set the fit to zero for A0V by subtracting this value from the fit. 
The color of a star 2/3 of the way from A0V to A1V (i.e., $V - K_S$ = 0.041, as should be roughly appropriate for Vega as shown below) is then $K_S - W1 = 0.0010 \pm 0.0020$. 

We made an independent determination by estimating $K - L$ based on linear fits on the values from A0V through F0V from a number of standard infrared color compilations, with the A0V colors forced to zero. The results are 0.0024, 0.0017, and 0.0013 respectively from 
 \citet{bessell1988},  \citet{koornneef1983}, and \citet{mamajek2018}. The slightly higher values are consistent with there being a number of rapid rotators among the  average A stars for which these colors pertain, resulting in a ``cooler'' SED than implied by the optical spectrum. 
 
 We adopt a value of 0.002 for the observed intrinsic $K_S$ - W1 color of Vega.  We used this value along with those for $H-K_S$ and $J-K_S$ derived above  to correct the values for J, H, and K to the equivalent values normalized to the 3.5 $\mu$m region. 
The adjusted measurements of Vega relative to Sirius are the ones given in Table 3.

The agreement for all the comparisons in Table 3 is excellent with an rms scatter of only 0.015 mag. However, the summary also shows that the results from \citet{cohen1992} may be smaller by about 0.005 mag than the rest,  as also suggested by \citet{bohlin2014}.  This difference is within the errors but suggests caution in interpreting these results. Another issue is that the \citet{cohen1992} values represent the comparison of theoretical SEDs normalized to a smaller number of absolutely calibrated measurements. To avoid over-weighting them in the average, we give them the weight of two measurements (given that there are two filter sets) although nine values are reported. We then find an average value of 1.370 magnitudes for the difference between the two stars near 3.5 $\mu$m. This value depends on the assumed $K-L$ color, since it offsets the values at the shorter wavelengths. However, given the weighting we have been using, adopting the difference between the observed value and the one from SED models affects the derived difference by only $\sim$ 0.001 mag. 

\begin{deluxetable}{lcccc}
\tabletypesize{\scriptsize}
\tablecaption{Comparison of Vega and Sirius in the Near Infrared}
\tablewidth{0pt}
\tablehead{
\colhead {Band\tablenotemark{a}} &
\colhead {Vega $F_\nu$ (Jy)} &
\colhead {Sirius $F_\nu$ (Jy)}  &
\colhead {difference\tablenotemark{b} (mag.)}  &
\colhead {reference} 
}

\startdata
$J_n$\tablenotemark{c}  &  1575.3  &	  5688.3  &  1.368	&   \citet{cohen1992}   \\
$K_n$\tablenotemark{c}	& 640.1  &  2262.1	&  1.369	&  \citet{cohen1992}  \\
$L_n$\tablenotemark{c}	&  246.0  &  860.9	&	1.360	&  \citet{cohen1992}  \\
$J$	& 1631.0 	&  5896.4  &	1.369	&   \citet{cohen1992}  \\
$H$	&  1049.7	&  3730.6    &	1.366	&  \citet{cohen1992} \\
$K$	&  655.0	&  2315.2	&	1.369	&  \citet{cohen1992}  \\
$L$	&  276.4	&  967.8	&	1.359	&   \citet{cohen1992} \\
$L'$	&  248.1	&  868.4	&	1.361	&  \citet{cohen1992}  \\
$M$	&  159.7	&  557.5	&	1.357	&   \citet{cohen1992}  \\
$[3.6]$  & --	 & --	&  	1.38	&   \citet{marengo2009}  \\
$[4.5] $  & --	&  --	&  	1.38	&   \citet{marengo2009} \\
$[5.8]  $  & --  &  --	&  	1.38	&   \citet{marengo2009}  \\
MSX $B1  $  & --  &  --      &  1.342	&   \citet{price2004}  \\
MSX $B2$   & --  &  ---      &  1.363	&   \citet{price2004}  \\
MSX $A$   & --  &  ---      &  1.358	&   \citet{price2004}  \\
DIRBE\tablenotemark{d}  &  --  &  --  &   1.395   &  \citet{smith2004}     \\
\hline
average\tablenotemark{a} & -- & --  &  1.370     &  ---  \\
 rms scatter  & --  &  --  &  0.012  & --  \\
\enddata

\tablenotetext{a}{Only bands that do not extend past 11 $\mu$m are included, to avoid any excess around Vega. Specifically, the MSX $A$ band is not significantly affected by the far infrared excess component \citep{su2013}.}
\tablenotetext{b} {$J$ band difference is reduced by 0.026 mag, $H$ band by 0.011 mag, and $K_S$ by 0.002 mag to account for the temperature gradient over the surface of Vega and the hot dust excess. }
\tablenotetext{c} {Narrowband filters centered roughly on the indicated standard bands. See \citet{selby1988, cohen1992}.}
\tablenotetext{d} {Because of the relatively large errors quoted for the DIRBE results, we show the weighted average of the three 2.2 - 4.9$\mu$m measurements; it has an indicated rms eror of 0.16 mag and is dominated by the 2.2 $\mu$m measurement.}

\end{deluxetable}

\subsubsection{Determining the $K_S$ Magnitude of Vega}

 The photometry listed by \citet{johnson1966} includes 178 measurements of Vega at $K$, but with some significant outliers. We have obtained a ``best'' value by averaging the results of two calculations. In the first, we fitted a Gaussian to the full distribution, obtaining a centroid at  $-$ 0.0178. The width of the gaussian implies a 1-$\sigma$ error of 0.035 per measurement, so the nominal error is $<$ 0.003. In the second, we rejected high and low outliers that fell more than 0.07 away from the mean (i.e., 2 $\sigma$ from the Gaussian fit), and averaged the remaining measurements to obtain $-$0.0161 with an rms scatter of 0.032, i.e., implying an uncertainty in the average $<$ 0.003. We adopt $-0.017 \pm 0.004$.  This becomes a $K_S$-band magnitude of $-$0.024 on the transformed 2MASS scale (using the transformation in Appendix {\bf A}). This value is equivalent within the errors to the finding by \citet{maiz2018} that the Vega-based \citep{cohen2003} $K_S$ zero point in 2MASS is $-0.015 \pm 0.005$. 
 
 A total $K_S$ of -0.024, of which 0.013 mag is contributed by hot dust \citep{absil2006, absil2008}, is consistent with other parameters of the star. The intrinsic $K_S$ magnitude of Vega (i.e., without the hot dust excess) is -0.011 and its intrinsic $V - K_S$ = 0.041. 
 This color puts it 2/3 of the way from A0V to A1V, where we compare with the colors tabulated by \citet{mamajek2018} but force zero color for A0V stars by subtracting the values quoted for them from the rest. These values are consistent with expectations for the color with the temperature gradient over the surface of the star, lending credibility to the $K_S$ magnitude.

\subsubsection{Putting Sirius on the 2MASS Scale}

	The conversion of the magnitude difference between Vega and Sirius to a magnitude for the latter star is influenced by the $1.26 \pm 0.27 \pm 0.06$\% excess for Vega from emission by hot dust \citep{absil2006, 
absil2008}, where the first error describes the statistical uncertainty and the second is a possible systematic error due to uncertainties in the stellar photospheric model. The inability to detect this excess at 10 $\mu$m \citep{liu2004, ertel2018} might suggest that it is variable. However, the uniformity of the results listed in Table 3 argues against this possibility. We have assumed that the excess is constant over the time of the measurements listed in this table. Our comparison has also assumed that the fractional excess above the photospheric emission is wavelength-independent from 2 to 7 $\mu$m. Models for the emission of the hot dust for two suggested compositions, carbon and FeO \citep{rieke2016, kirchschlager2017}, indicate that this assumption is probably justified, although carbon dust might produce a spectrum somewhat redder than Rayleigh Jeans and FeO somewhat bluer. The variability and color behavior of the emission by the hot dust adds to the uncertainty in the final Sirius magnitude.

We found that Sirius is 1.370 mag brighter than Vega, which in turn is at $K_S$ = -0.024, or -0.026 at IRAC Band 1, applying our nominal correction for the temperature gradient on Vega. We therefore deduce a magnitude of $ -1.396 \pm 0.010$ for Sirius at IRAC Band 1 and the same value at $K_S$ on the basis that the magnitude should be wavelength-independent.

\subsection{Use of {\it Spitzer} Photometry on Bright Field Stars}

Yet another approach is to use the photometry from \citet{su2021} of selected field stars relative to Sirius in the IRAC Band 1, see row 4 of Figure 1. The approach is summarized in Table~4. It proceeds by using a standard color difference for the spectral type of the star to correct the 2MASS $K_S$ magnitude\footnote{That is, the value obtained from transformation of heritage photometry.} to the magnitude at IRAC Band 1, and then applying the magnitude difference from Sirius to obtain an estimate of the magnitude of the latter star. Since we are assuming that Sirius has zero color difference in $K_S$ - IRAC1, the same value applies at $K_S$. 

The foundation of the approach in this section is that photometry of 1\% accuracy can be obtained using the wings of the Spitzer PSF \citep{marengo2009}. 
The accuracy of the PSF wing-fitting photometry is uniquely possible with Spitzer because of its extremely stable beryllium optics, stemming from operation in an isothermal environment at a temperature where the coefficient of expansion for beryllium is virtually zero. The IRAC instrument has also proven to be very stable \citep{carey2008, carey2012, lowrance2014}. However, the complex faint structures of the IRAC PSF are very sensitive to the dither pattern in which the data were taken. We therefore designed specific observations to utilize this technique. In an accompanying paper, \citet{su2021} reduce and discuss this photometry, including tests that support the achievement of measurements with errors $\lesssim$ 1\%. 
 They tabulate average values for the magnitude difference of each star relative to Sirius; we use these values in our analysis. 

We have to apply the $K_S$ - IRAC1 color difference appropriate for the spectral type to convert these measurements to $K_S$.  We have derived these color differences for main sequence stars (hence not for the K0III star $\beta$ Gem) as described in Appendix {\bf B}, finding


$$
K_S - IRAC1   =  -0.001228 \times  x^4 +0.014382 \times x^3 - 0.042158 \times  x^2 + 0.057587 \times  x,
\eqno(1)
$$

\noindent
where $x$ = $V - K_S$. To obtain IRAC Band 1 magnitudes for comparison with the saturated photometry with Spitzer, we have applied this term to the transformed $K_S$ magnitudes for all of the stars measured by \citet{su2021} except HD 19476\footnote{The two available K-band measurements for this star differ by 0.06 mag; a review suggests that the measurement by \citet{alonso1994} is correct, but since it is a single measurement it does not deserve equal weight with the others. We show in \citet{su2021} that its $K_S$ is nonetheless consistent with the results for the other, higher-weight stars.}, which we exclude because only a single measurement is available to determine its $K_S$ magnitude.  Averaging the results for all the other stars, we obtain an estimate of the $K_S$ magnitude of Sirius of -1.394. 

As discussed in Appendix {\bf B}, there are two possible small biases to this result, the offset of $0.0014 \pm 0.0021$ magnitudes in comparing the transformed $K_S$ magnitudes with the 2MASS ones, and a similar possible offset due to extinction of the stars used to determine equation (1). Both of these biases work to result in making our magnitude for Sirius being too negative, perhaps by as much as 0.004 if we apply both. This full possible offset would affect our final value for Sirius combining all of our approaches by $\lesssim$ 0.001 mag.  


\begin{deluxetable}{lcccccccc}
\label{spitzermethod}
\tabletypesize{\footnotesize}
\tablecaption{$K_S$ magnitude of Sirius Using Spitzer Photometric Transfer}
\tablewidth{0pt}
\tablehead{
\colhead {Star} &
\colhead {Spec. } &
\colhead{V\tablenotemark{a}}   &
\colhead {$K_S$} &
\colhead{refs.\tablenotemark{b}}   &
\colhead{$K_S$-} &
\colhead {IRAC1} &
\colhead{Sirius } &
\colhead{Sirius } \\
\colhead {} &
\colhead {Type} &
\colhead {}    &
\colhead {mag} &
\colhead{}   &
\colhead{IRAC1\tablenotemark{c}} &
\colhead {mag} &
\colhead{Transfer} &
\colhead{mag} 
} 
\startdata
HD 19373  &  G0V  & 4.05 & 2.650  &1,2,3,4 &  0.033   &  2.617  & 4.001  &  -1.384  \\
HD 30652 & F6V    & 3.19 & 2.076   & 1,4,5,6,7  & 0.030   & 2.046  &   3.440  &  -1.394  \\
HD 61421 &  F5IV  & 0.37 & -0.677  & 1,4,6,8,9  & 0.029  & -0.706 &   0.671  &  -1.377  \\
HD 62509 &  K0III  &1.14  & -1.118  & 1,3,4,8,9  & 0.07  &  -1.188  &  0.202  &  -1.390  \\
HD 102870 &  F9V  & 3.60 & 2.293  & 1,7,5,10 & 0.032  &  2.261  &  3.652 &  -1.391  \\
HD 126660 &  F7V  & 4.05 & 2.808  & 4,5,11 & 0.031 &  2.777  &  4.184  &  -1.407  \\
HD 142860 &  F6V  & 3.84 & 2.623  & 1,4,5,7 & 0.031   &  2.592  &   3.991  &  -1.399  \\
HD 173667 &  F5.5IV-V  & 4.19  & 3.039  & 3,4,5,7 &  0.030  & 3.009  &   4.410  &  -1.401  \\
HD 215648 &  F6V  & 4.20 & 2.900\tablenotemark{d}  & 3,4,12 & 0.032  &  2.849\tablenotemark{d}  &  4.233  &  -1.389  \\
\hline
mean (average)  &  &  &  &  &  &  &  &   -1.393  \\
rms scatter  &  &  &  &   &  &  &  &  0.009  \\
standard error   & &  & &  &  &  &  &  \\
~~~of the mean & &  & & &   &  &  & $<$0.004 \\
\enddata
\tablenotetext{a}{from \citet{stassun2019}}
\tablenotetext{b} {1 \citet{johnson1966}; 2 \citet{kidger2003}; 3 \citet{selby1988}; 4 this work (relative to $K_S$ for HD 165459, with color correction as in Appendix B); 5 \citet{aumann1991}; 6 \citet{glass1974}; 7 \citet{casagrande2014}; 8 \citet{ukirt2018}; 9 \citet{smith2004}; 10 \citet{alonso1994}; 11 \citet{milone2005}; 12 \citet{alonso1998}. }
\tablenotetext{c}{Based on $K_S$ magnitude and expected color difference from equation (1) (see also Appendix A),  except for $\beta$ Gem = HD 62509, where we have taken the color difference from \citet{tokunaga2000}. }
\tablenotetext{d}{There is a star 11\farcs1 from HD 215648 \citep{raghavan10} with a $K_S$ magnitude from 2MASS of 7.30. This star would have been outside the photometric aperture for references 3 and 12, but would have been included in reference 4. The $K_S$ magnitude tabulated here is corrected for the influence of this additional star. For the transfer to Sirius, however, we have added the contribution of the nearby star (0.019 mag) to the brightness of HD 215648, since both stars would have been measured together. The combined magnitude is listed in column 7 as the ``IRAC1 mag.'' }
\end{deluxetable}

\subsection{Direct Transfer between Sirius and HD 165459}

The preceding section relied on a direct transfer from Sirius to stars with accurate $K_S$ magnitudes transformed from other systems. We can also utilize a direct transfer from Sirius to HD 165459, a star that is exceptionally well measured with Spitzer and is within the range for {\it unsaturated} 2MASS measurements. We show this approach in the  extreme right column of Figure 1; we will use HD 165459 as a transfer standard from a suite of 2MASS A-star measurements to Sirius.  This allows a determination of the $K_S$ magnitude of Sirius without the need to transform any photometry into the 2MASS system. 

HD165459 has been classified as a A3V star \citep{grenier1999} and to have a $T_{eff} \sim 8565$K  \citep{bohlin2017}, also corresponding to a spectral type of A3V. The IRAC data demonstrate that the star shows no variability, as is expected for a main sequence A-star. Its 2MASS survey $K_S$ magnitude is 6.584 $\pm$ 0.027, and its V magnitude (converted from Tycho) is 6.86, i.e., V $-$ K$_S$ = 0.28 $\pm$ 0.03, very similar to the intrinsic color for A3V, 0.25 \citep{pecaut2013, mamajek2018}, indicating it is only very lightly reddened (although reddening is not important for our use of it as a transfer standard).  In confirmation, a value of $E(B-V) = 0.021 - 0.023$ was found by \citet{bohlin2017}. It has a debris disk infrared excess, by a factor of $\sim$ 1.5 at 24$\mu$m and of 14 relative to the stellar photosphere at 70$\mu$m \citep{su2006}, but again this will have no effect on our use of it as a transfer standard in the IRAC band 1 at 3.6 $\mu$m, or even as a primary standard at wavelengths short of $\sim$ 5 $\mu$m.

First, we consider the saturated Spitzer photometry of HD 165459.  If we take all six numbers for IRAC Band 1 from \citet{su2021}, that is all SNRs and means and medians and average we get a ratio of 1552, rms scatter of 4. If we do the same with Band 2, the average factor is 1564 $\pm$ 4, and if we take all eight determinations (both bands) and average all of them we get 1558 $\pm$ 7.  This corresponds to 7.981 mag as the difference between the two stars. 

We discuss our derivation of an accurate $K_S$ magnitude for HD 165459 in Section~\ref{faint}.  The resulting average determination of the $K_S$ magnitude is 6.588 $\pm$ 0.007. This value agrees nearly perfectly with the 2MASS measurement of 6.584 $\pm$ 0.027, but with a much smaller error. Applying the transfer to Sirius, we obtain a $K_S$ magnitude for it of -1.389 based on the IRAC Band 1 transfer. If we go through the same steps with the IRAC Band 2 photometry from Su et al., we find a magnitude for Sirius of -1.398, and if we average both determinations we obtain the best value from this approach of  -1.393. This value is virtually identical to the final average of all the determinations (see next section).Thus, if we eliminate it from the average, we obtain the nearly the same estimate of the magnitude of HD 165459, 6.586, this time utilizing the measurements of Sirius that are independent of HD 165459 and the Spitzer transfer from Sirius to HD 165459. The two determinations, one from using the Spiter photometry to determine a high accuracy $K_S$ magnitude and the other to transfer directly from Sirius, are virtually identical. Since the are completely independent, we will adopt their average, 6.587.

\subsection{Summary, Final Value for $K_S$ of Sirius}

Table~5 collects the five determinations of the magnitude of Sirius. They are in excellent agreement, indicating that there is no significant systematic error for any of them. We can therefore average them to obtain a ``best'' estimate of the magnitude of the star, -1.395. To estimate the error, we took all the individual measurements within each of the five approaches and computed the rms scatter among the 46 such measurements. The value is 0.009 mag, again indicative of high accuracy throughout this study. A reasonable final error is the $\pm$ peak-to-peak error, i.e., half of the full peak-to-peak error. That is 0.005 mag.

\begin{deluxetable}{lc}
\label{summary}
\tabletypesize{\footnotesize}
\tablecaption{Summary, 2MASS $K_S$ Magnitude of Sirius}
\tablewidth{0pt}
\tablehead{
\colhead {Approach} &
\colhead {$K_S$ (2MASS)}
} 
\startdata
{transformed heritage photometry}  &  -1.3944  \\
{using DIRBE photometry}  &  -1.4018  \\
{via comparison with Vega}  &  -1.394  \\
{direct transfer to field stars}  & -1.3934  \\
{direct transfer Sirius to HD 165459}  &  -1.3915  \\
\hline
average  &  -1.3950  \\
rms scatter  &  0.009\tablenotemark{a}  \\
indicated error of average  & $<$ 0.005\tablenotemark{b}\\
\enddata
\tablenotetext{a} {This scatter is the standard deviation for all the individual determinations within each method, a total of 46.}
\tablenotetext{b} {We assume an upper limit to the error to be half of the peak-to-peak range of the five results.} 
\end{deluxetable}

\section{Transferring to Fainter Stars}
\label{faint}

In this section, we transfer the result for Sirius to fainter stars, starting with all the A-stars from \citet{krick2021}. We then emphasize two stars that have been used extensively for previous standard star networks: (1) BD +60 1753 is a well-behaved 9.6$^{th}$ magnitude  A1V star that, despite its distance of $\sim$ 500 pc (Gaia DR2), is very lightly reddened \citep{bohlin2017} -- it was a primary standard for {\it Spitzer} and hence has more than 400 measurements in IRAC Bands 1 and 2 extending throughout the cold and warm missions; (2) 2MASS J16313382+3008465 = GSPC P330-E was used as the primary solar-type calibrator for NICMOS and has about 60 measurements with Spitzer in IRAC Bands 1 and 2 -- it is of type G2V and has about 0.1 $-$ 0.15 magnitudes of extinction ($A_V$) \citep{bohlin2017}.  

\subsection{Accurate $K_S$ magnitudes for A-Stars}

 \citet{krick2021} published accurate Spitzer photometry for 18 stars with types between B9IV and A6V, providing the opportunity to derive very accurate $K_S$ magnitudes for this list.  To obtain an estimate of the $K_S$ magnitude of a target star independent of the actual 2MASS measurement of that star, we proceed as follows. We select another of the A-type stars, calculate its magnitude difference to the target star in IRAC Band 1 corrected for the expected $K_S$ $-$ IRAC1 color difference from equation (1), and adjust the $K_S$ {\it observed} by 2MASS for this star by this amount.  We repeat this for all the other A-type stars that are not significantly obscured (see next paragraph) and average all of the resulting estimates of the $K_S$ magnitude of the target star to get a much more accurate value. These values can be found in Table~ 6. The scatter in the individual magnitudes for a given target star is 2.2\%, consistent with the expected scatter in the 2MASS magnitudes (i.e., the normalization according to the IRAC measurements has not added significant error). The indicated net uncertainty for the average for a given target star is $\sim$ 0.7\%. 

An issue for this approach is that equation (1) is only applicable for unobscured or very lightly obscured stars. For four of the stars, comparison of the V $-$ $K_S$ color with expectations for their spectral type suggests $A_V >$ 0.2: HD 2811, HD 163466, 2MJ17571324+6703409, and 2MJ18120957+6329423.  As discussed further in Appendix {\bf B}, in these cases equation (1) will not return a highly accurate result. Given the uncertainties in the spectral types, it is not feasible to separate the stellar color and the extinction accurately, so we have left these four stars out of the determinations of averaged accurate $K_S$ magnitudes. Of course, this is not an obstacle in determining such magnitudes for {\it them}. 

\begin{deluxetable}{lcccc}
\label{summary}
\tabletypesize{\footnotesize}
\tablecaption{Accurate $K_S$ Magnitudes for Calibration Sample A-Stars}
\tablewidth{0pt}
\tablehead{
\colhead {Star} &
\colhead {Type}  &
\colhead{reference\tablenotemark{a}} &
\colhead {$\left< K_S \right>$\tablenotemark{b}}  &
\colhead {$K_S$ (2MASS)}  
} 
\startdata
HD 2811  &  A3V   &  1    &   7.056     &  7.057     $\pm$ 0.024     \\
HD 14943  & A5V & 1   & 5.434  & 5.439  $\pm$ 0.018 \\
HD 37725  & A3V & 2   & 7.913  & 7.902   $\pm$ 0.018 \\
eta01Dor  & A0V & 3  & 5.764  & 5.751   $\pm$ 0.024 \\
HD 55677  & A2V & 4   & 9.161  & 9.156   $\pm$ 0.021 \\
HD 116405  & B9IV & 5  & 8.489  & 8.476  $\pm$ 0.020 \\
HR 5467  & A1V & 6 & 5.811  & 5.756   $\pm$ 0.018 \\
BD 60 1753 & A1V & 7  & 9.639  & 9.645   $\pm$ 0.015 \\
HD 158485  & A3V  & 6 & 6.134  & 6.145  $\pm$ 0.023 \\
HD 166205  & A0V  & 6 & 4.239  & 4.258   $\pm$ 0.029 \\
2MJ17325264+7104431 & A4V & 7 & 12.246  & 12.254   $\pm$ 0.030 \\
HD 163466  & A5V & 7 & 6.346  & 6.339   $\pm$ 0.018 \\
2MJ17571324+6703409  & A4V & 7  & 11.163  & 11.155   $\pm$ 0.023 \\
2MJ18022716+6043356  & A2V  & 7 & 11.854  & 11.832   $\pm$ 0.018 \\
HD 165459  & A3V & 5  & 6.588  & 6.584   $\pm$ 0.027 \\
2MJ18083474+6927286 & A6V &  7 & 11.584  & 11.532   $\pm$ 0.019 \\
2MJ18120957+6329423  & A3V & 2  & 11.277  & 11.286  $\pm$ 0.019 \\
HD 180609  & A0V & 2 & 9.103  & 9.117   $\pm$ 0.019 \\
\enddata
\tablenotetext{a} {References for spectral types: (1) \citet{houk1978}; (2) \citet{reach2005}; (3) \citet{houk1975}; (4) \citet{fehrenbach1966}; (5) \citet{grenier1999}; (6) \citet{abt1995}; (7) \citet{bohlin2017}}
\tablenotetext{b} {High precision magnitude (see text); indicated error 0.007 mag}
\end{deluxetable}

\subsection{Transfer from HD 165459 to BD +60 1753 and GSPC P330-E}

As already pointed out, our direct transfer from Sirius to HD 165459 yields a nearly identical $K_S$ magnitude for the latter star as we obtain for the high-precision $K_S$ magnitude in Table 6. We averaged the results of these two approaches to derive a magnitude of $K_S$ = 6.587, which we use as the reference for transferring Sirius to fainter stars. We now obtain brightnesses relative to HD 165459 for fainter stars to continue this process.

We have used four approaches to obtain an accurate transfer from HD 165459 to BD +60 1753, summarized in Table~7.  Since the two stars are both of early A-type and with low levels of extinction, we have ignored color terms and extinction corrections given that we are working reasonably far into the infrared. The first two entries in the table are just based on the flux ratios in \citet{krick2021}; in Figure 1, the two bands are merged into a single item in row 6. The third entry, row 7 in Figure 1, is from the high-precision $K_S$ magnitudes in Table~6.

 The fourth entry, in the right-hand column in Figure 1,  is based on an automated pipeline comparing the measurements of HD 165459 to BD +60 1753 directly, as discussed in Section 2. Based on this reduction, we compared HD 165459 and BD +60 1753 in two ways. In the first, we fitted the signal for each star with a linear dependence with time (to allow for the slow decrease in sensitivity, see, e.g., \citet{krick2021}), rejecting a few obvious outliers, and calculated the average of the ratio of the fits. This approach makes use of all the data. In the second, we took the ratio of the signals from the two stars whenever they were both measured within a 24 hour interval, so time variations in sensitivity are irrelevant. This method uses about 70\% of the data. The two determinations agree well, indicating a magnitude difference of 3.049 in IRAC Band 1. The value for Band 2 is identical to within $\sim$ 0.2\%. Although in principle the error in this difference should be very small, there is about a 0.7\% difference in the value between cold and warm  Spitzer missions (perhaps due to residual errors in the linearity corrections), so we assign an error of 0.005 for our measurements that combine in roughly equal measure data from the cold and warm mission segments. We base this estimate on the measurements that the nonlinearities for both mission segments were corrected to better than 1\% up to saturation \citep{lowrance2014}.

\begin{deluxetable}{lcccc}
\label{BDxfer}
\tabletypesize{\footnotesize}
\tablecaption{$K_S$ Magnitudes of BD +60 1753 and GSPC P330-E}
\tablewidth{0pt}
\tablehead{
\colhead {Approach} &
\colhead {HD 165459 to} &
\colhead {BD +60 1753} &
\colhead {BD to P330} &
\colhead {P330-E} \\
\colhead {} &
\colhead { BD +60 1753 difference} &
\colhead {$K_S$ magnitude} &
\colhead{difference} &
\colhead {$K_S$ magnitude}\\
\colhead {} &
\colhead {magnitudes} &
\colhead{} &
\colhead {magnitudes} &
\colhead{} 
} 
\startdata
\citet{krick2021} I1  &  3.043  & 9.630\tablenotemark{a}  & 1.771\tablenotemark{b} &  11.441 \\
\citet{krick2021} I2  &  3.046  & 9.633\tablenotemark{a}  & --\tablenotemark{c} &  --  \\
high precision $K_S$ (Table 6)  &  3.051\tablenotemark{d}  & 9.638  & -- &  --  \\
Direct IRAC comparison  &  3.049  & 9.636\tablenotemark{a}  & -- &  --  \\
NEOWISE W1  &  --  & --  & 1.783\tablenotemark{e} &  11.433  \\
CALSPEC  &  -- & --  &  1.855\tablenotemark{f} &  11.424\tablenotemark{g} \\
2MASS cal star  & -- &  -- & -- & 11.424 \\
\citet{persson1998}  & --  &  --  &  --  &  11.418  \\
\citet{leggett2006}  &  --  &  --  &  --  &  11.430 \\  
\hline
averages  &  --  &  9.634  &  --  &  11.428  \\
rms scatter  &  --  &  0.004  &  --  &  0.009  \\
\enddata
\tablenotetext{a} {Applying IRAC magnitude difference to high precision $K_S$ from Table 6}
\tablenotetext{b}{IRAC1 value. Corrected for P330-E magnitude by 0.035 for $K_S$ - IRAC1 color and 0.006 for extinction \citep{gordon2021}.}
\tablenotetext{c}{Not used for P330-E because of uncertainty in depth of CO fundamental.}
\tablenotetext{d}{Not used for magnitude of BD +60 1753, but tabulated for completeness.}
\tablenotetext{e} {Relative to average for BD +60 1753, 9.632. Corrected for P330-E magnitude by 0.014 for $K_S$ - W1 color and 0.004 for extinction.}
\tablenotetext{f} {Difference in CALSPEC values at $H$-band, relative to average for BD +60 1753, 9.632. } 
\tablenotetext{g} {Corrected for P330-E by 0.056 for H $-$ $K_S$ color and 0.007 for extinction.}
\end{deluxetable}

We now turn to the G2V calibration star, GSPC P330-E, as shown in the bottom four rows in Figure 1 and again with results summarized in Table~7. 
One route to estimate the $K_S$ magnitude of GSPC P330-E is to use the NEOWISE post-cryogenic mission \citep{mainzer2011}  data (unfortunately, HD 165459 is above the saturation limit for NEOWISE, so its photometry is compromised) and apply the difference between BD +60 1753 and HD 165459 derived above. There are about 150 measurements of GSPC P330-E and nearly 400 of BD +60 1753 in the NEOWISE single-frame photometry at the Infrared Science Archive (IRSA). We have rejected the high and low outliers (1 $-$ 2\% of the total measurements) and then computed mean magnitudes finding a difference of $1.783 \pm 0.003$ magnitudes. 

For a second estimate, we obtained a magnitude difference at H-band for BD +60 1753 and GSPC P330-E from the Wide Field Camera 3 (WFC3) IR grism spectra in  \citet{bohlin2019}. To do so, we averaged the ratio of the spectra of BD +60 1753 and GSPC P330-E from 1.59 to 1.72 $\mu$m (i.e., to the cutoff long wavelength of the grism spectra and from a short wavelength limit so the ratio is centered on the effective wavelength of the H band). The result is entered in column 4 of Table 7. To convert it to a value at $K_S$, we took the H $-$ $K_S$ value (0.076) for a G2V star from \citet{casagrande2012} but corrected this value by 0.020 mag for the zero point offset for 2MASS H $-$ $K_S$. This correction is based on determinations of 0.019 mag \citep{rieke2008}, 0.026 (Cutri, cited in \citet{rieke2008}) and 0.019 \citep{maiz2018}. 

P-330E was a prime 2MASS calibration star \citep{skrutskie2006} and thus was observed many times during the course of the survey. These results provide a third estimate of its $K_S$ magnitude. The result is entered into Table 7; it should have a very small nominal error. 

Finally, \citet{persson1998} conducted a campaign of high accuracy photometry of solar-type stars in support of the calibration of NICMOS on HST.  They obtained two measurements of GSPC P330-E in slightly different K bands, $K(Persson) = 11.419 \pm 0.007$ and $K_S (Persson) = 11.429 \pm 0.006$. However, these measurements were made when the 2MASS calibration was still in a formative stage. We use the 2MASS measurements of the entire set measured by \citet{persson1998} to derive linear transformations as a function of $J - K$ to obtain corrections of $K (Persson) - K_S(2MASS)$ = 0.003 and $K_S(Persson) - K_S (2MASS)$ = 0.010, yielding values of $K_S$ of 11.416 and 11.419 for GSPC P330-E respectively; we adopt 11.418.  \citet{leggett2006} also report accurate groundbased photometry of the star, obtaining a $K_{MKO}$ magnitude of 11.419, which converts to a $K_S$ magnitude of 11.430 using the transformation given in that paper. 

\subsection{Comparison with CALSPEC}

The good behavior of BD +60 1753 in the models of \citet{bohlin2017} is encouraging that an independent infrared calibration can be derived from the CALSPEC models, which are normalized in the visible. To do so, we carried out synthetic photometry in the 2MASS $K_S$ band for all the CALSPEC stars, with the relative spectral response in the $K_S$ band from \citet{cohen2003} and on the system with the zero point determined in that work. We find  $K_S$ magnitudes of -1.383 and 9.637 respectively for Sirius and BD +60 1753. The value relative to Sirius when it is assigned $K_S$ = -1.395  (thus consistent with our definitions) is 9.625 for BD +60 1753, within 1\% of the value we have derived in Table ~7. This agreement shows that the infrared- and optically-based calibrations are in reasonable agreement. This comparison will be discussed further in a later paper in this series where we review the infrared absolute calibration data. 

\section{Extending an accurate calibration over the entire sky}

As always, extending an accurate calibration over the entire sky is a challenge. The 2MASS survey is one option. A signal to noise ratio of $\sim$ 100 (as needed for a 1\% calibration reference) was achieved for sources brighter than $K_S \sim$ 12 but ``For 8.5 $< K_s <$ 13 default uncertainties
are consistently in the range 0.02 $-$ 0.03 mag. This limiting profile-fit uncertainty is attributable to PSF mismatch in the profile-fit algorithm, driven by undersampling due to the coarse 2\farcs0 pixel size.'' - \citet{skrutskie2006}. In addition, there are possibly large-scale non-uniformities up to the 2\% level, see Figure 18 of \citet{skrutskie2006}. Many of these issues are mitigated for the 35 2MASS ``tracer'' fields \citep{nikolaev2000}, which include many of the stars in the \citet{persson1998} list and by association on the sky, fainter field stars. \citet{leggett2020} provide a listing of 81 equatorial standards that are much fainter (median $K_S$ = 17.5), and whose magnitudes are traceable to the 2MASS system through the UKIRT Infrared Deep Sky
Survey (UKIDSS) \citep{hodgkin2009} and Visible and Infrared Survey Telescope for Astronomy (VISTA) \citep{gonzalez2018} parent surveys. 

An interesting alternative utilizes the NEOWISE data as obtained from the CatWISE2020\citep{eisen2020} catalog \citep{marocco2021}. We have used the Spitzer photometry from \citet{krick2021} to perform a test. We used the measurements of stars with IRAC Band 1 flux densities from 3.56 to 192.51 mJy, corresponding to W1 magnitudes of 12.24 to 7.95. The signal to noise and ability to identify background sources that might compromise the photometry both dictate the fainter limit, and the brighter one corresponds to the CatWISE W1 saturation limit. The stars for this test are listed in Table 8. We compared their magnitudes with those from \citet{krick2021} by taking the following steps. First, we averaged the magnitudes in W1 and W2 (for GSPC P330-E, we made a 0.04 mag color correction in W2 before averaging).   We assigned this average magnitude to a pseudo-IRAC 1 Band (which is intermediate in wavelength) and ratioed the actual IRAC Band 1 flux to the flux in the pseudo-band, adjusting the zero point of the pseudo-band to set the average of all the ratios to 1.  The resulting zero point, 282.9 Jy, is in good agreement with that from \citet{reach2005}, 280.0 $\pm$ 4.1 Jy for the real IRAC band. Because the CatWISE2020 values are based on a huge number of measurements of each star, our hope was that the resulting magnitudes would have small internal errors (any overall offset due to calibration differences is removed by our adjustment of the zero point). In fact, Table 8 shows that the results agree with the IRAC Band 1 measurements to within about 1\%. In making use of the CatWISE2020 data to this degree of accuracy, one should check the images to be sure that there are no potentially confusing sources, even to very faint levels, and we found two cases where this was a possibility. Removing them reduces the rms scatter to just below 1\%. Although the sample size is small, these results suggest that the CatWISE2020 data have substantial potential as an all-sky standard network. 

Applying this network to the conventional $K$ band requires a color difference correction; however, we have found that there are significant discrepancies among different estimates for these corrections, particularly outside the $0.7 < V-K_S < 1.7$ range. 
 We have compared three color difference studies:  (1) the tabulation of standard colors from \citet{pecaut2013}\footnote{available at  \url{http://www.pas.rochester.edu/$\sim$emamajek/EEM\_dwarf\_UBVIJHK\_colors\_Teff.txt}}; (2) the photometric colors derived by \citet{jian2017, deng2020}; and (3) colors based on SDSS-classified stars by  \citet{davenport2014}. As shown in Figure 2, expressed as a function of $V - K_S$ color, there are discrepancies of 1 $-$ 2\% (with a limited range of better agreement between  the values of \citet{pecaut2013} and \citet{jian2017} for $0.7 \le V-K_S \le 1.7$). However, for 1 $\lesssim$ $V - K_S$ $<\lesssim$ 2, the slopes of the corrections are in reasonably good agreement, indicating that the primary cause of the discrepancies at least for this range is the normalization. Although the all-sky CatWISE photometry is promising for many applications, it appears that to exploit it to high accuracy requires more accurate transformations to other photometric systems. Fortunately, the CatWISE measurements can still be used to full accuracy in applications where such transfers are not necessary (e.g., in our applications in this paper). 

\begin{figure*}
\epsscale{.8}
\plotone{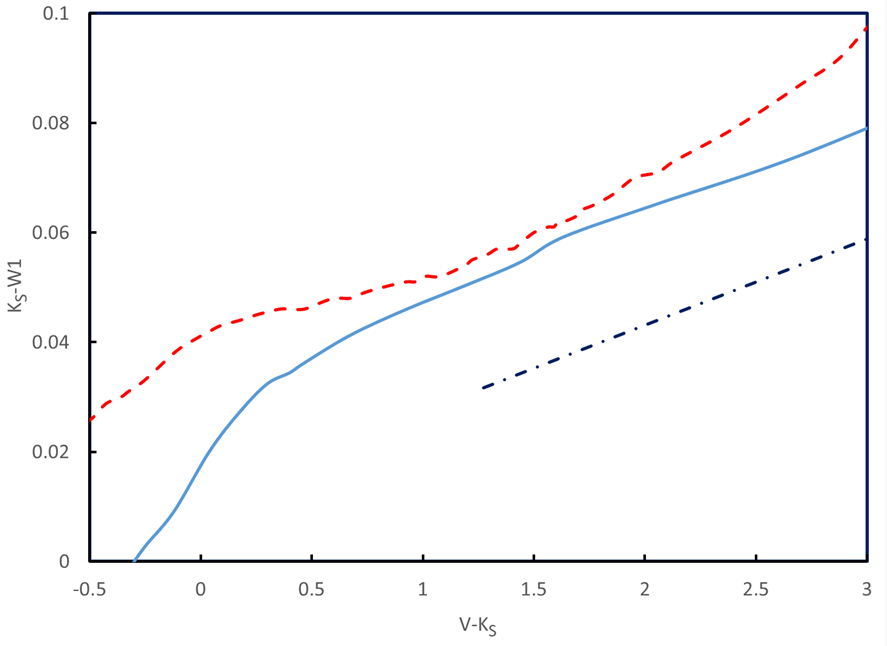}
\caption{Comparison of proposed color difference $K_S - W1$ as a function of  $V - K_S$ for main sequence stars. The dashed line is from \citet{mamajek2018}, the solid one from \citet{jian2017,deng2020}, and the dash-dot one from \citet{davenport2014}.
}
\label{flow}
\end{figure*}

\begin{deluxetable}{lcccc}
\tabletypesize{\scriptsize}
\tablecaption{Test of CatWISE2020 Photometry}
\tablewidth{0pt}
\tablehead{
\colhead {Star} &
\colhead {IRAC1 (mJy)} &
\colhead {Average WISE magnitude}  &
\colhead { Flux Density (mJy)}  &
\colhead {Ratio}  \\
\colhead {} &
\colhead {} &
\colhead {}  &
\colhead { for WISE average mag\tablenotemark{a}}  &
\colhead {IRAC1/$<$W1+W2$>$}  
}
\startdata
HD 37725  &  192.51  &	7.934  &  189.68 & 1.0149 \\
HD 55677  &  61.03  &	9.162  & 61.21 & 0.9970 \\
HD 116405  &  113.31  &	8.500  &  112.57 & 1.0066 \\
P330-E  &  7.74  &	11.387  &  7.886 & 0.9815\tablenotemark{b} \\
BD +60 1753 &  39.56  &	9.643  &  39.304 & 1.0065 \\
2M J17325264+7104431  &  3.56  &	12.250  &  3.560 & 1.00004 \\
2M J17571324+6703409  &  9.65  &	11.176  &  9.577 & 1.0076\tablenotemark{c} \\
2M J18022716+6043356  &  5.11  &	11.866  &  5.075 & 1.0069 \\
2M J18083474+6927286	  &  6.55  &	11.565  &  6.696 & 0.9782\tablenotemark{c} \\
2M J18120957+6329423	  & 8.69  &	11.281  &  8.694 & 0.9995 \\
HD 180609  &  64.38  &	9.109  &  64.244  &  1.0021 \\
\hline
rms scatter\tablenotemark{d}  &  --  &  --  & --  &  0.0112  \\ 
rms scatter outliers rejected\tablenotemark{e} & --  &  --  &  --  &  0.0092  \\
\enddata
\tablenotetext{a}{Assuming a zero point of 282.9 Jy for this pseudo-band.}
\tablenotetext{b}{Includes allowance for W1 - W2 = -0.04}
\tablenotetext{c}{Frame image shows nearby clutter, photometry may have larger than typical errors}
\tablenotetext{d} {For all the stars}
\tablenotetext{e} {With 2M J17571324+6703409 and 2M J18083474+6927286 eliminated }
\end{deluxetable}

\section{Summary}

We have used multiple approaches to determine an accurate $K_S$ magnitude for Sirius (-1.395), HD 165459 (6.587), BD +60 1753 (9.634), and GSPC P330-E (11.428) on the 2MASS scale. This work will improve the transfer of the accurate physical calibration of Sirius (e.g., \citet{price2004}, to be reviewed in a future paper) to the other two stars, which are widely used standard calibration stars and are central to the absolute calibration of JWST. In turn, the absolute calibration can be transferred to much fainter $K_S$ measurements \citep[e.g.,][]{leggett2020}, or possibly over the entire sky using either $K_S$ measurements or NEOWISE ones.

\section{Acknowledgements}

We thank Micaela Bagley and Rafia Bushra for extensive work on data reduction and analysis. Ralph Bohlin provided very helpful comments. The paper is based on observations made with the Spitzer Space Telescope, which was operated by the Jet Propulsion Laboratory, California Institute of Technology. This publication also makes use of data products from the NearEarth Object Wide-field Infrared Survey Explorer (NEOWISE), which
is a project of the Jet Propulsion Laboratory/California Institute of Technology. NEOWISE is funded by the National Aeronautics and
Space Administration.  We used data from the Two Micron All Sky Survey (2MASS), a joint project of the University of Massachusetts and the Infrared Processing and Analysis Center/California Institute of Technology, funded by the National Aeronautics and Space Administration and the National Science Foundation. This work was supported by NASA grants NNX13AD82G and 1255094.

\facility{Spitzer, WISE, 2MASS, COBE}

\eject

\eject

\appendix{\bf Appendix A: Transformations and Standard Colors}

This appendix summarizes the transformed (to 2MASS) magnitudes of the stars used in our study, and indicates where the transformations were derived. 
Many transformations were obtained from \citet{carpenter2001}, which were used for the photometry from  \citet{feast1990}, \citet{fluks1994}, \citet{bouchet1991}, \citet{glass1974}\footnote{Except for those used in the direct photometry of Sirius, see Section 3.1.}, \& UKIRT starred magnitudes. The photometry of \citet{casagrande2012} was transformed as in \citet{koen2007}. Transformations were derived for this work for the remaining sets of photometry. Doing so is challenging for the \citet{johnson1966} photometry because there is very little overlap between it and non-saturated 2MASS photometry. We proceeded by transforming to intermediate sets of stars with reasonable overlap with both the Johnson and 2MASS results, as summarized in Table 9.

\begin{deluxetable}{lcc}[h!]
\tabletypesize{\footnotesize}
\tablecaption{Transformations from Johnson $K$ to 2MASS $K_S$ Magnitudes}
\tablewidth{0pt}
\tablehead{
\colhead {Method} &
\colhead {Transformation}  &
\colhead {Number\tablenotemark{a}}
} 
\startdata
via ESO standards\tablenotemark{b}  & $K_S\left(2MASS \right)=K\left( Johnson \right)-0.055 + 0.028(J-K)$  &  97 \\
via SAAO\tablenotemark{c} &  $K_S\left(2MASS \right)=K\left( Johnson \right)-0.038+0.025(J-K)$ & 82 \\
via Kidger \& Mart\'in-Luis\tablenotemark{d} &  $K_S\left(2MASS \right)=K\left( Johnson \right)-0.033+0.024(J-K)$  & 39 \\
via UKIRT bright standards\tablenotemark{e} &  $K_S\left(2MASS \right)=K\left( Johnson \right)- 0.038 + 0.022 (J-K)$ &  54 \\
\hline
adopted   &   $K_S\left(2MASS \right)=K\left( Johnson \right)- 0.041 + 0.024 (J-K)$  & --  \\
\enddata
\tablenotetext{a}{the number of stars in common with those in \citet{johnson1966} on which the transformation is based - virtually all these stars 
are main sequence A through K types}
\tablenotetext{b}{\citet{bouchet1991}}
\tablenotetext{c}{\citet{carter1990}}
\tablenotetext{d}{\citet{kidger2003}}
\tablenotetext{e}{\citet{ukirt2018}}
\end{deluxetable}

The derivation of transformations for the photometry of \citet{aumann1991}, \citet{alonso1998} (Telescopio Carlos S\'anchez (TCS) system), \citet{kidger2003} and \citet{milone2005}  was relatively straightforward and they are shown in the equations below.

$$
J_{2MASS} = J_{TCS} - 0.004  + 0.049 \times (J - K)_{TCS}
$$
$$
H_{2MASS} = H_{TCS} + 0.040 - 0.030 \times (J - K)_{TCS}
$$
$$
K_{2MASS} = K_{TCS} + 0.003 - 0.007 \times (J - K)_{TCS}
$$

$$
J_{2MASS} = J_{Aumann} -0.039 + 0.069 \times (J - K)_{Aumann}
$$
$$
H_{2MASS} = H_{Aumann} + 0.015 + 0.001 \times (J - K)_{Aumann}
$$
$$
K_{2MASS} = K_{Aumann} -0.019 + 0.001 \times (J - K)_{Aumann}
$$
$$
J_{2MASS} = J_{Kidger}-0.0199+0.0483 \times (J - K)_{Kidger}
$$
$$
H_{2MASS} = H_{Kidger} + 0.01083  - 0.01563 \times (J - K)_{Kidger}
$$
$$
K_{2MASS} = K_{Kidger} + 0.00081 + 0.00257 \times (J - K)_{Kidger}
$$
$$
K_{2MASS} = K_{Milone} - 7.3398 - 0.0979\times(J-K)_{Milone} + 0.2657\times(J-K)^2_{Milone}
$$

 The magnitudes used in this work are the averages of all the available photometry transformed into the 2MASS system. The values and references are provided in the Table 10 below. 
 
 \begin{deluxetable}{ccccccccc}[h!]
\label{transphot}
\tabletypesize{\footnotesize}
\tablecaption{Transformed Photometry\tablenotemark{a}}
\tablewidth{0pt}
\tablehead{
\colhead {star} &
\colhead {V}  &
\colhead{IRAC1}  &
\colhead {J}  &
\colhead{H}  &
\colhead{$K_S$}  &
\colhead{Refs.\tablenotemark{b}}  &
\colhead{IRAC 1} &
\colhead{$K_S$ }  \\
\colhead {} &
\colhead {}  &
\colhead{(Jy)}  &
\colhead {}  &
\colhead{}  &
\colhead{}  &
\colhead{}  &
\colhead{(mag\tablenotemark{c})} &
\colhead{heritage\tablenotemark{d}} 
} 
\startdata
HD000142	&	5.700	&	4.630	&	4.757	&	4.533	&	4.464	&	1,2	&	4.440	&	4.453	\\						
HD000739	&	5.237	&	6.409	&	4.401	&	4.201	&	4.139	&	2	&	4.087	&	4.139	\\						
HD001835	&	6.383	&	3.222	&	5.231	&	4.950	&	4.856	&	1,3,4,5	&	4.834	&	4.856	\\						
HD003795	&	6.131	&	5.482	&	4.780	&	4.400	&	4.318	&	1,2	&	4.257	&	4.303	\\						
HD004307	&	6.147	&	4.022	&	4.981	&	4.723	&	4.610	&	1,6	&	4.593	&	4.599	\\						
HD003302	&	5.509	&	4.803	&	4.739	&	4.522	&	4.479	&	1,2	&	4.400	&	4.455	\\				
\enddata

\tablenotetext{a}{Table 10 is published in its entirety in machine-readable format.
      A portion is shown here for guidance regarding its form and content.}

\tablenotetext{b} {1, 2MASS; 2, Su \& Rieke, unpublished; 3, \citet{allen1983}; 4, \citet{bouchet1991};  5, \citet{alonso1998}; 6,
 \citet{kidger2003}; 7, \citet{aumann1991}; 8, \citet{carter1990}; 9, \citet{glass1974} ; 10, \citet{mcgregor1994}; 
 11, \citet{johnson1966}; 12, \citet{groote1983}}
 
 \tablenotetext{c}{Zero point taken to be 276.5 Jy to force zero color for V $-$ $K_S$ = 0.}
 
 \tablenotetext{d} {$K_S$ magnitude based only on heritage photometry. The JHK magnitudes in the previous columns include 2MASS photometry as well as the transformed heritage photometry whenever unsaturated 2MASS photometry is available.}
 
\end{deluxetable}

We also calculated specific transformations for the determination of the magnitude of Sirius by conventional photometry, as discussed in Section 2.1:

$$
K_{2MASS} = K_{Glass} - 0.0243 - 0.0057 \times (J - K)_{Glass}
$$

\noindent
This provided a magnitude of -1.375 for Sirius; if we had used the \citet{carpenter2001} transformation, we would have obtained -1.373, indicating that our caution in deriving a new transformation from the \citet{glass1974} photometry itself made little difference. The transformation for the \citet{engels1981} photometry is

$$
K_{2MASS} = K_{Engels} -0.0396 - 0.0018 \times (J - K) - 0.0202 \times (J - K)^2_{Engels}
$$

\eject

\appendix{\bf Appendix B: $K_S$ to IRAC Band 1 Color Behavior}

\begin{figure}[h!]
\epsscale{.8}
\plotone{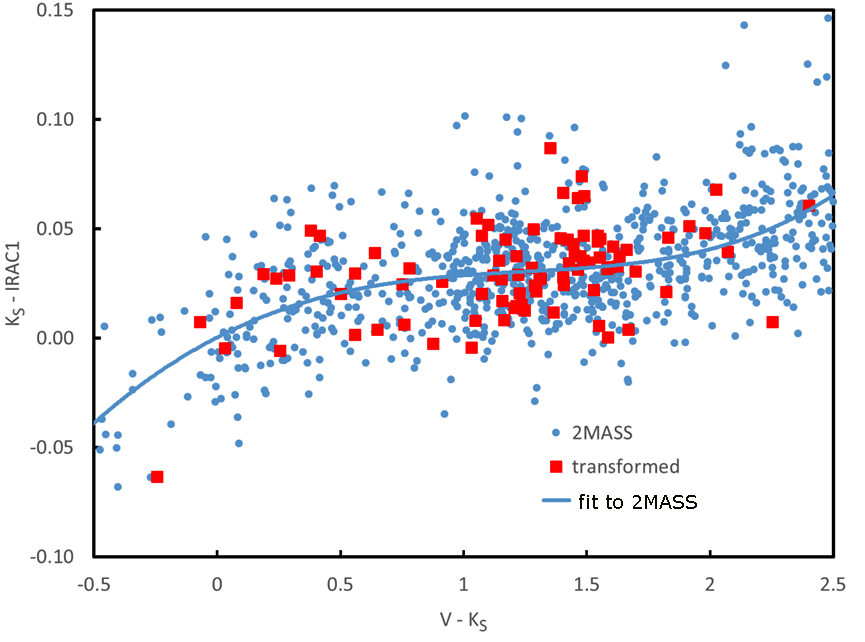}
\caption{Color-color plot showing $K_S - IRAC1$ vs. $V - K_S$ for stars from {\it Spitzer}  PID 70076 with  $K_S$ measurements transformed from heritage $K$ photometry (red squares, 89 stars) compared with stars with only 2MASS $K_S$ measurements (small blue dots, 820 stars), over the range of $V - K_S$ represented by the former sample, $-0.5 < V-K_S < 2.5$. The blue line is a 4$^{th}$ order polynomial fitted to the blue dots.  
}
\label{cc}
\end{figure}

Figure~\ref{cc} compares the $K_S$ photometry transformed from heritage measurements with the 2MASS photometry, based on the $K_S - IRAC1$ color as a function of $V - K_S$. (The full set of transformed photometry is provided in Table 10 and the IRAC Band 1 photometry in Table 11.) The underlying assumption is that the IRAC photometry is stable, as found in multiple studies \citep[e.g.,][]{krick2021}. Although for best values, we merge

\begin{deluxetable}{ccc}[h!]
\label{transphot}
\tabletypesize{\footnotesize}
\tablecaption{Photometry in IRAC Band 1\tablenotemark{a}}
\tablewidth{0pt}
\tablehead{
\colhead {star} &
\colhead {Band 1 flux (Jy)}  &
\colhead{Error (Jy)}  \\
} 
\startdata
HD000319	&	1.817	&	0.050	\\
HD001237	&	3.303	&	0.040	\\
HD003302	&	4.803	&	0.049	\\
HD003369	&	3.683	&	0.049	\\
HD003823	&	4.498	&	0.052	\\
HD004150	&	5.413	&	0.046	\\
HD004277	&	0.866	&	0.029	\\
HD004391	&	5.515	&	0.056	\\
HD006569	&	0.338	&	0.005	\\
HD006767	&	3.500	&	0.050	\\
HD007590	&	2.424	&	0.032	\\
HD008224	&	1.518	&	0.030	\\
HD008556	&	3.149	&	0.049	\\
\enddata
\tablenotetext{a}{Table 11 is published in its entirety in machine-readable format.
      A portion is shown here for guidance regarding its form and content.}
\end{deluxetable}

2MASS measurements when they are available with transformed data, in this case we have separated the two and plot the results {\it only} for the heritage $K_S$ values.  The figure also shows similar color-color measurements for 2MASS measurements compared with IRAC Band 1. The blue line is a 4$^{th}$ order polynomial fitted to the 2MASS-only values.  We have adjusted the $K_S - IRAC1$ color to be zero at $V-K_S$ = 0 by making a small adjustment in the assumed calibration of the IRAC photometry (this affects only a constant in the fit, which is effectively set to zero). We assume an error for the IRAC measurements of 1\% \citep{hora2008} and for both the transformed and 2MASS-only $K_S$ ones of 2\%, in the former case as demonstrated in Appendix C and in the latter, a typical value even for bright stars \citep{skrutskie2006}. These errors are combined quadratically and then reduced $\chi^2$ values calculated for each of the heritage and 2MASS-only photometry. The values are 0.68 and 0.96 respectively for the transformed and native 2MASS sets of measurements. The average offset in $K_S - IRAC1$ between the transformed photometry and the line fitted to the 2MASS-only results is $0.0014 \pm 0.0021$ magnitudes, where the error is the standard deviation.

As with Appendix C, this test confirms the accuracy of the transformed $K_S$ photometry, in this case on a different sample of (much fainter) stars. It further demonstrates that the transformed values are accurately on the 2MASS $K_S$ photometric system.  Therefore, we merge the two sets of data and determine an overall relation between $V - K_S$ and $K_S - IRAC1$:

$$
K_S - IRAC1   =  -0.001228 \times  x^4 +0.014382 \times x^3 - 0.042158 \times  x^2 + 0.057587 \times  x,
\eqno(2)
$$

\noindent
where $x$ = $V - K_S$.

The limiting faint $K_S$ magnitude for this sample is $\sim$ 6, except for a small number of very red and low luminosity stars. Taking values of $M_K$ from \citet{mamajek2018}, the stars of type A0V and later therefore lie within 100 pc, i.e., are within the Local Bubble. Similarly stars with $V - K_S >$ 1 (later than F4V) lie within 60 pc. Within 100 pc, the level of extinction is uniformly very low, with $E(b-y) < 0.03$ \citep{reis2011}, i.e., $A_V < 0.1$ and typically $\sim$ 0.04. Low levels of extinction move the points in Figure~\ref{cc} diagonally, and values of $A_V < 0.1$ (or $\sim$ 0.04)have very little overall effect on the relation in equation (2) (based on the extinction relations in \citet{rieke1985} and \citet{gordon2021} - ($A_V$ = 0.04 corresponds to E($K_S - IRAC1) \sim 0.002$)). We can therefore use this equation as a general conversion for IRAC Band 1 measurements to $K_S$ for main-sequence stars.

\appendix{\bf Appendix C: Test of the Accuracy of GG and Transformed Groundbased Photometry}

To test the accuracy of both the COBE DIRBE 2.2 $\mu$m photometry (converted to K magnitudes with an assigned zero point) and our transformed heritage photometry onto the 2MASS $K_S$ system, we have compared K-magnitudes for all the stars brighter than 2$^{nd}$ magnitude. This gives us an initial sample of 507 stars, which we have trimmed to eliminate stars that are variable, or have missing data or nearby stars bright enough to affect the DIRBE photometry. For the fainter stars, the DIRBE photometry is not sufficiently sensitive to eliminate variability to the level of 3\%, which is the threshold we have adopted. We therefore have also cut stars of types where variability is common but the DIRBE data are not definitive in this regard. The overall statistics are summarized in Table 12. We fitted a linear transformation in $J - K$ vs. $2.2_{DIRBE}$ $-$ $K_S$ and corrected the DIRBE results to be equivalent to the 2MASS system. 

\begin{deluxetable}{cccccc}[h!]
\label{sampstat}
\tabletypesize{\footnotesize}
\tablecaption{Source Counts for Transformed and DIRBE Photometry}
\tablewidth{0pt}
\tablehead{
\colhead {Total} &
\colhead {DIRBE variables\tablenotemark{a}}  &
\colhead{Missing Data}  &
\colhead {Nearby Stars\tablenotemark{b}}  &
\colhead{C, S, or MIII\tablenotemark{c}}  &
\colhead{Final Sample} \\
} 
\startdata
507 & 53  & 63  & 44 & 64 & 283  \\
\enddata
\tablenotetext{a}{As determined by \citet{smith2004}.}
\tablenotetext{a}{We rejected all cases where the 2MASS survey showed a star at $K_S$ $\gtrsim$ 3\% as bright as the target star within the DIRBE beam.}
\tablenotetext{b}{Where the DIRBE data had inadequate signal to noise to probe variability to the 1$\sigma \sim$ 2\% level, we rejected all stars of types S, C, and MIII since they have a relative high probability of being variable. }

\end{deluxetable}

\begin{deluxetable}{ccccccc}[h!]
\label{errors}
\tabletypesize{\footnotesize}
\tablecaption{Error Analysis of Transformed and DIRBE Photometry}
\tablewidth{0pt}
\tablehead{
\colhead {Error Source} &
\colhead {rms}  &
\colhead{rms}  &
\colhead {rms}  &
\colhead{rms}  &
\colhead{rms}  &
\colhead{rms} \\
} 
\startdata
DIRBE assumed & 0  & 0.005  & 0.010 & 0.015 & 0.020 & 0.022  \\
2MASS derived & 0.022  & 0.0215  & 0.019 & 0.0155 & 0.008 & 0  \\
2MASS 2$\sigma$ upper limit & 0.027  & 0.026  & 0.024 & 0.021 & 0.017 & --  \\
\enddata
\end{deluxetable}

We found that the rms scatter in $2.2_{DIRBE}$ $-$ $K_S$ for the 127 stars brighter than 1$^{st}$ magnitude is 0.026 mag. The rms scatter for the full sample is larger, 0.043 mag, reflecting the lower signal to noise in the DIRBE measurements of the fainter stars. We therefore turned to a $\chi^2$  analysis of the full sample of 283 stars. We assumed that the uncertainty for each star consisted of the $rms$ combination of $<$err$>$ tabulated for that star in \citet{smith2004}, plus a value for additional DIRBE and transformed 2MASS photometry that we varied to understand the permissible tradeoffs. The results are in Table 13. They show that the $rms$ errors in our transformed photometry are no more than $\sim$ 2\%, similar to the errors in the 2MASS photometry of non-saturated stars \citep{skrutskie2006}. From evaluating the scatter in standard colors for the stars with transformed photometry, we believe that the derived values significantly $<$ 2\% are overly optimistic. Thus, with care to eliminate stars in the DIRBE beam, this instrument can deliver photometry at the 1\% level, according to this study.

\end{document}